\documentclass{article}

\usepackage{upgreek}
\usepackage{setspace}
\usepackage{graphicx}
\usepackage{amssymb}
\usepackage{subcaption}
\usepackage{tikz}
\usepackage[font=footnotesize,labelfont=bf]{caption}
\usepackage{amsmath}
\usepackage{multicol}
\usepackage{hyperref}\hypersetup{colorlinks = true, citecolor = blue, linkcolor=blue}

\textheight \dimexpr \pdfpageheight -2\topmargin -1.9in

\begin{document}
\title{\vspace{-5.6cm}\textbf{\hspace*{-1.93cm}{\mbox{{\Large Three-dimensional simulations of spatiotemporal instabilities in}}} {\Large a magneto-optical trap}}}
\author{
\hspace*{-1.7cm}{\normalsize M. Gaudesius$^{1,2,\footnote{\hspace{0cm}Corresponding author: gd.mar900@gmail.com}}$ , Y.-C. Zhang$^{3,4}$, T. Pohl$^3$, R. Kaiser$^1$, and G. Labeyrie$^1$} \\
\hspace*{-1.6cm}\vspace{-0.1cm}{\small\textit{ $^{1}$Universit\'{e} C\^{o}te d'Azur, CNRS, Institut de Physique de Nice, 06560 Valbonne, France }} \\ 
\hspace*{-1.3cm}\vspace{-0.1cm}{\small\textit{ $^{2}$Department of Physics and Astronomy, University of Oklahoma, Norman, Oklahoma 73019, USA }} \\
\hspace*{-1.7cm}\vspace{-0.1cm}{\small\textit{ $^{3}$Center for Complex Quantum Systems, Department of Physics and Astronomy, }} \\
\hspace*{-1.6cm}\vspace{-0.1cm}{\small\textit{ Aarhus University, DK-8000 Aarhus C, Denmark }} \\
\hspace*{-1.2cm}\vspace{-0.1cm}{\small\textit{ $^{4}$MOE Key Laboratory for Nonequilibrium Synthesis and Modulation of Condensed Matter, }} \\
\hspace*{-1.2cm}\vspace{-0.1cm}{\small\textit{ Shaanxi Key Laboratory of Quantum Information and Quantum Optoelectronic Devices, }} \\
\hspace*{-1.2cm}\vspace{-0.1cm}{\small\textit{ School of Physics, Xi'an Jiaotong University, Xi'an 710049, People's Republic of China }}
 }
\date{}
\maketitle

\vspace{-0.25cm}
\leftskip=-1.15cm\rightskip=-1.15cm
{\small Large clouds of atoms in a magneto-optical trap (MOT) are known to exhibit spatiotemporal instabilities when the frequency of the trapping lasers comes close to the atomic resonance. Such instabilities possess similarities with stars and confined plasmas, where corresponding nonlinearities may give rise to spontaneous oscillations. In this paper, we describe the kinetic model that has recently been employed in three-dimensional (3D) simulations of spatiotemporal instabilities in a MOT, yielding qualitative agreements with experimentally observed instability thresholds and regimes. Details surrounding its implementation are included, and the impact of its physical effects on the instabilities is investigated to improve the understanding of the complex mechanism at work.}

\vspace{10pt}
\begin{multicols}{2}\setlength{\columnsep}{2pt}

\leftskip=-0.5cm
\section{\hspace{-0.38cm}{Introduction}}\label{sec:1}

\leftskip=-3cm\rightskip=0.15cm
A magneto-optical trap (MOT) is nowadays extensively used to study properties of neutral atoms, or as a part of more elaborate setups for creating, e.g., quantum degenerate gases. The limit of a large atom number $N$, in particular, has attracted interest in various applications, notably random lasing \cite{0:RandomLasing}, Anderson localization \cite{0:AndersonLocalization}, self-organization \cite{0:SelfOrganization}, superradiance \cite{0:Superradiance} and subradiance \cite{0:Subradiance}. This limit can, moreover, allow for investigations of nonlinear phenomena that possess similarities with pulsating stars \cite{3:stellar, 4:stellar} and unstable plasmas \cite{5:plasma}-\cite{7:plasma}. 

At low $N$, a MOT is governed by single-atom physics, such that each atom in the cloud is independently subjected to a cooling and confining force applied by the laser beams in the presence of the magnetic field. At large $N$, many-atom physics become important as $\textit{collective}$ forces appear, e.g., the shadow force \cite{10:ShadowForce} and the rescattering force \cite{11:Wieman}. The shadow force is caused by an imbalance of beam intensities in the cloud due to attenuation, which occurs because of light's scattering as it traverses through the cloud. This force is compressive but is countered by the repulsive, Coulomb-like force caused by photon rescattering between the atoms. With these antagonistic forces present, large MOT clouds can exhibit spatiotemporal instabilities in the form of spontaneous oscillations.

Spatiotemporal instabilities have been studied in several MOT configurations \cite{3:stellar, 11:Wieman}-\cite{13:PM1}, and many theoretical models have been explored to gain insight into the experimentally observed features \cite{25:Pohl}-\cite{29:PM2}. In our case, the so-called \textit{balanced} MOT configuration is considered, where the laser beams are independent and have the same, constant intensities before entering the cloud. In our recent works \cite{1:MG} and \cite{2:MG}, we have made experimental observations of instability thresholds and regimes, respectively, and reached qualitative\;\;agreements\;\;with\;\;results\;\;of\;\;our

\leftskip=0.15cm\rightskip=-3cm 
\noindent three-dimensional (3D) simulations. In the present work, we explain the kinetic model and the technicalities employed in these simulations. Moreover, the simulations are employed to investigate the role of different effects on the instabilities.

The remainder of this article is structured as follows. In Sec. \hyperref[sec:a]{II.A}, the kinetic model employed in the 3D simulations is described, followed up by Sec. \hyperref[sec:b]{II.B}, where details surrounding its implementation are covered. Then, Sec. \hyperref[sec:3]{III} investigates the impact of its physical effects on the instabilities. Finally, Sec. \hyperref[sec:4]{IV} provides a conclusion and a discussion on the future perspectives. 

\leftskip=1.12cm
\section{\hspace{-0.38cm}{Simulation method and \\ \hspace*{0.37cm}{approximations} }}\label{sec:2}
\vspace{-6pt}
\leftskip=2.82cm
\subsection{\hspace{-0.30cm}{Kinetic model}}\label{sec:a}
\vspace{2pt}

\begin{figure*}
\vspace*{-102pt}
\hspace*{-69pt}\includegraphics[scale=0.5]{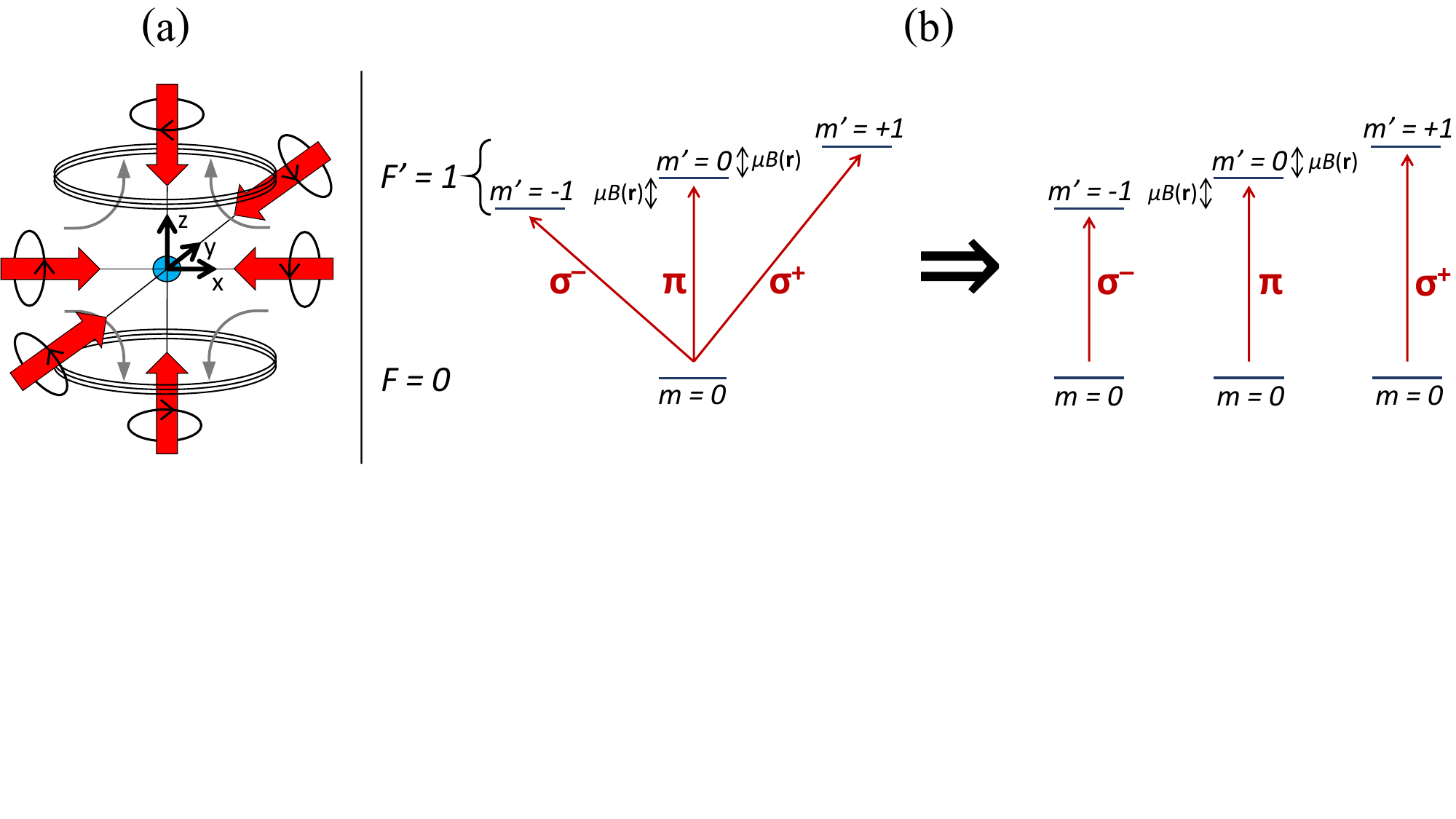}
\vspace*{-130pt}
 \captionsetup{width=1.49\linewidth}
  \caption{(a) Sketch of a MOT, displaying the considered conventions for the MOT magnetic field and beams in the derivation of the kinetic model employed in our 3D simulations of instabilities. The coils produce a magnetic field (curved arrows) that has a positive gradient along the $\mathsf{z}$ axis. Accordingly, the beams have the following circular helicities (circular arrows): Right-handed for the $\mathsf{z}$-axis beams, and left-handed for the $\mathsf{x}$- and $\mathsf{y}$-axis beams. (b) Simplification of the Zeeman sub-level structure of the hyperfine transition $F=0\rightarrow F'=1$, in the employed kinetic model. Each of the three Zeeman transitions $m=0\rightarrow m'=-1,0,+1$ between $F=0$ and $F'=1$ is treated as an independent 2-level system (right picture). They are induced by, respectively, only $\sigma^-$, $\pi$, $\sigma^+$ polarized light. The MOT magnetic field leads to the Zeeman splitting of the hyperfine levels depending on the atom position [$\mu B(\textbf{r})$].}
\label{fig:1}
\end{figure*}

\leftskip=0.15cm\rightskip=-3cm 
\noindent In the development of our kinetic model, the considered conventions for the MOT magnetic field and beams are displayed in Fig. \ref{fig:1}(a). The magnetic field, which is quadrupole in nature, is assumed to be $\textbf{B}(\textbf{r})=B'\left(-\frac{x}{2},-\frac{y}{2},z\right)$, where $B'>0$ is the field gradient along the $\mathsf{z}$ axis, and $\textbf{r}=(x,y,z)$ is the atom position with the origin at the trap center. This assumption holds when $|\textbf{r}|$ is much smaller than both the radius of the MOT coils and the separation between them. In accordance with the chosen magnetic field convention, the $\mathsf{z}$-axis beams have right-handed circular helicity, whereas the $\mathsf{x}$- and $\mathsf{y}$-axis beams have left-handed circular helicity.

We base our model on the hyperfine transition $F=0 \rightarrow F'=1$. It is evidently much simpler than in our experiments, $F=2 \rightarrow F'=3$ (totaling 12 Zeeman sublevels) \cite{1:MG, 2:MG}, but allows for a proper description of the features related to the magnetic field and\;\;light\;\;polariza-

\leftskip=-3cm\rightskip=0.15cm

\noindent tion. As depicted in Fig. \ref{fig:1}(b), each of the three Zeeman transitions $m = 0 \rightarrow m' = -1, 0,+1$ between the hyperfine levels is treated as an independent 2-level system and driven by, respectively, $\sigma^-$, $\pi$, $\sigma^+$ polarized light. We expect this approximation to hold only in the regime of low saturation \cite{extra:Rotation}, $s=\frac{I_{\infty}/I_{sat}}{1+\frac{4\Delta^2}{\Gamma^2}}\ll1$, where $I_{\infty}$ is the intensity of a beam before entering the cloud, $I_{sat}$ is the saturation intensity, $\Delta=\omega_L - \omega_0$ is the detuning of the laser frequency $\omega_L$ from the atomic transition $m = 0 \rightarrow m' = 0$ frequency $\omega_0$, and $\Gamma$ is the natural linewidth. 

We also neglect sub-Doppler effects, which seems to be reasonable considering that large MOT clouds have been observed to be mostly unaffected by these effects \cite{extra1:sub-Doppler}. 

We include the following main four physical effects in our model: (i) the mean cooling and confining force stemming from the MOT beams, hereafter referred to as the trapping force; (ii) the diffusion resulting from its fluctuations; (iii) the beam intensity attenuation caused by the light's scattering in the cloud; and (iv) the rescattering force due to photon exchange between the atoms. In the following, we explain how these effects are described.

\leftskip=-0.15cm
\subsubsection{\hspace{-0.21cm}Trapping force}\label{sec:a_i}

\leftskip=-3cm\rightskip=0.15cm
To describe the trapping force, we use the standard Doppler model. It relies on an assumption, expected to be valid for $s\ll1$, that this force can be expressed as a sum of six radiation pressure forces, one for each beam:

\begin{equation}
\label{eq:0}
\begin{aligned}
\hspace*{-85pt}
\textbf{F}_{tr}(\textbf{r},\textbf{v})= \sum_{\alpha=\mathsf{x},\mathsf{y},\mathsf{z}}\textbf{F}_{\alpha}^{+}(\textbf{r},\textbf{v}) + \textbf{F}_{\alpha}^{-}(\textbf{r},\textbf{v})
\end{aligned}
\end{equation}

\noindent where $\textbf{v}=(v_\mathsf{x},v_\mathsf{y},v_\mathsf{z})$ is the atom velocity, and $\textbf{F}_{\alpha}^{+}$ and $\textbf{F}_{\alpha}^{-}$ are the radiation pressure forces exerted by the beams traveling in, respectively, positive and negative directions of $\alpha = \mathsf{x},\mathsf{y},\mathsf{z}$, given by

\begin{equation}
\label{eq:1}
\begin{aligned}
\hspace*{-85pt}
\textbf{F}_{\alpha}^{\pm}(\textbf{r},\textbf{v})=\textbf{F}_{\alpha,-}^{\pm}(\textbf{r},\textbf{v})+\textbf{F}_{\alpha,0}^{\pm}(\textbf{r},\textbf{v})+\textbf{F}_{\alpha,+}^{\pm}(\textbf{r},\textbf{v})
\end{aligned}
\end{equation}
\noindent where $\textbf{F}_{\alpha,-}^{\pm}$, $\textbf{F}_{\alpha,0}^{\pm}$, $\textbf{F}_{\alpha,+}^{\pm}$ are the radiation pressure forces acting on the atom's\;\;2-level\;\;transitions\;\;that\;\;are\;\;driven

\leftskip=0.15cm\rightskip=-3cm 
\noindent by, respectively, $\sigma^-$, $\pi$, $\sigma^+$ polarized light, given by
\begin{equation}
\label{eq:2}
\begin{aligned}
\hspace{20pt}
\textbf{F}_{\alpha,q}^{\pm}(\textbf{r},\textbf{v}) = \pm\frac{p_{\alpha,q}^{\pm}(\textbf{r}) I_{\alpha}^{\pm}(\textbf{r},\textbf{v}) \sigma_{\alpha,q}^{\pm}(\textbf{r},\textbf{v}) }{c}\boldsymbol{\hat{\alpha}}
\end{aligned}
\end{equation}

\noindent In this equation, $q=-,0,+$ refers to the respective $\sigma^-$, $\pi$, $\sigma^+$ transitions, $c$ is the vacuum light speed, and the remaining quantities are defined as follows. 

The coefficient $p_{\alpha,q}^\pm$ denotes the fraction of the positive or negative ($\pm$) $\hat{\boldsymbol{\alpha}}=\hat{\boldsymbol{\mathsf{x}}},\hat{\boldsymbol{\mathsf{y}}},\hat{\boldsymbol{\mathsf{z}}}$ directed light that drives the $\sigma^-$, $\pi$, or $\sigma^+$ ($q=-,0,$ or $+$) transition \cite{15:cadmium}. It introduces anisotropy to the trapping force. There is a total of 18 fractions, as there are three transitions corresponding to each of the six MOT beams: 
\begin{equation}
\label{eq:frac}
\begin{aligned}
\hspace*{15pt}
 p_{\alpha,q}^\pm(\textbf{r})=\begin{cases}
               \left( \frac{1}{2}\left[ 1 \pm \frac{\alpha'B'}{2B(\textbf{r})} \right] \right)^2\quad,\quad q = +\;(\sigma^+)\\
               \left( \frac{1}{2}\left[ 1 \mp \frac{\alpha'B'}{2B(\textbf{r})} \right] \right)^2\quad,\quad q = -\;(\sigma^-)\\
               1-(p_{\alpha,+}^\pm+p_{\alpha,-}^\pm)\quad,\quad q = 0\;\,\,(\pi)
            \end{cases}
\end{aligned}
\end{equation}

\noindent where $\alpha'=x,y,2z$ for, respectively, $\alpha = \mathsf{x},\mathsf{y},\mathsf{z}$, $B(\textbf{r})=B'\sqrt{z^2+\frac{1}{4}\left(x^2+y^2\right)}$ is the magnitude of the magnetic field $\textbf{B}(\textbf{r})$. The quantization axis has been chosen to be along the direction of $\textbf{B}(\textbf{r})$. Consequently, as shown in Fig. \ref{fig:1}(b), the Zeeman shifts of the excited levels with $m' = -1, 0,+1$ are given by $\mu_q(\textbf{r})= q\mu B(\textbf{r})$, where $q=-,0,+$, respectively, and $\mu$ is the gyromagnetic ratio. 

For the final quantities in Eq. (\ref{eq:2}), $I_{\mathsf{\alpha}}^{\pm}$ is the corresponding beam intensity, which is subject to attenuation as will be covered later, and $\sigma_{\alpha,q}^{\pm}$ is the corresponding scattering cross-section for a single 2-level transition of the atom, given by
\begin{equation}
\label{eq:3}
\begin{aligned}
\hspace*{15pt}
\sigma^{\pm}_{\alpha,q}(\textbf{r},\textbf{v})=\frac{\sigma_0}{1+\frac{I_{tot,q}(\textbf{r},\textbf{v})}{I_{sat}}+4\frac{(\Delta\mp k_Lv_{\alpha} - \mu_q(\textbf{r}))^2}{\Gamma^2}}
\end{aligned}
\end{equation}
\noindent where $\sigma_0=6\pi/k_L^2$ is the resonant scattering cross section, where $k_L=\omega_L/c$ denotes the laser wavenumber; $\mp k_Lv_{\alpha}$ is the\;\;Doppler shift for a positive or negative ($\pm$) beam;

\leftskip=-3cm\rightskip=0.15cm
\noindent $\mu_q$ is the Zeeman shift defined previously; and
\begin{equation}
\label{eq:4}
\hspace*{-89pt}
\begin{aligned}
I_{tot,q}(\textbf{r},\textbf{v})=\sum_{\alpha=\mathsf{x},\mathsf{y},\mathsf{z}}p_{\alpha,q}^{+}(\textbf{r}){I}^{+}_{\alpha}(\textbf{r},\textbf{v}) + p_{\alpha,q}^{-}(\textbf{r}){I}^{-}_{\alpha}(\textbf{r},\textbf{v})
\end{aligned}
\end{equation}

\noindent is the total beam intensity that a single 2-level transition receives, which is observed to be generally different for all three ($\sigma^-$, $\pi$, and $\sigma^+$) due to $p_{\alpha,q}^\pm$. Because this intensity enters into scattering cross sections, the beam cross-saturation effect gets naturally introduced into our model. In the section on attenuation, we mention a difficulty (luckily, solvable) that is encountered by having this effect included.

\vspace*{-8pt}
\leftskip=0.25cm
\subsubsection{\hspace{-0.21cm}Diffusion}\label{sec:a_ii}

\leftskip=-3cm\rightskip=0.15cm
The fluctuating part of the trapping force can be introduced via a momentum diffusion coefficient. Here we describe such diffusion processes approximately by adapting the known description for a 2-level atom in a single laser beam \cite{16:Tannoudji} to our 3D situation involving six laser beams. Explicitly, we write the following for the momentum diffusion coefficient:

\begin{equation}
\label{eq:5}
\hspace*{-75pt}
\begin{aligned}
D(\textbf{r},\textbf{v})=D_{vac}(\textbf{r},\textbf{v}) + D_{las}(\textbf{r},\textbf{v})
\end{aligned}
\end{equation}
\noindent where
\begin{equation}
\label{eq:5_vac}
\hspace*{-155pt}
\begin{aligned}
D_{vac}(\textbf{r},\textbf{v})=\hbar^2 k^2_L\frac{\Gamma}{4}\frac{s_{tot}(\textbf{r},\textbf{v})}{1+s_{tot}(\textbf{r},\textbf{v})}
\end{aligned}
\end{equation}
\begin{equation}
\label{eq:5_las}
\hspace*{-75pt}
\begin{aligned}
D_{las}(\textbf{r},\textbf{v})&=\hbar^2 k^2_L\frac{\Gamma}{4}\frac{s_{tot}(\textbf{r},\textbf{v})}{[1+s_{tot}(\textbf{r},\textbf{v})]^3}\\&\times\left\{1+\frac{12\Delta^2-\Gamma^2}{4\Delta^2+\Gamma^2}s_{tot}(\textbf{r},\textbf{v})+s_{tot}^2(\textbf{r},\textbf{v})\right\}
\end{aligned}
\end{equation}

\noindent are the momentum diffusion coefficients of, respectively, the vacuum and laser field, with $\hbar$ being the reduced Planck constant and $s_{tot}$ being the total saturation parameter that is a sum of the total saturation parameters $s_{tot,-}$, $s_{tot,0}$, $s_{tot,+}$ for the atom's 2-level transitions that are driven by, respectively, $\sigma^-$, $\pi$, $\sigma^+$ polarized light, i.e.,

\begin{equation}
\label{eq:stot}
\hspace*{-85pt}
\begin{aligned}
s_{tot}(\textbf{r},\textbf{v})=\sum_{q=-,0,+}s_{tot,q}(\textbf{r},\textbf{v})
\end{aligned}
\end{equation}
where

\begin{equation}
\label{eq:stot_q}
\hspace*{-85pt}
\begin{aligned}
s_{tot,q}(\textbf{r},\textbf{v})=\sum_{\alpha=\mathsf{x},\mathsf{y},\mathsf{z}}s^{+}_{\alpha,q}(\textbf{r},\textbf{v})+s^{-}_{\alpha,q}(\textbf{r},\textbf{v})
\end{aligned}
\end{equation}
and

\begin{equation}
\label{eq:salpha_q}
\hspace*{-85pt}
\begin{aligned}
s_{\alpha,q}^{\pm}(\textbf{r},\textbf{v})=\frac{p_{\alpha,q}^{\pm}(\textbf{r}){I}^{\pm}_{\alpha}(\textbf{r},\textbf{v})/I_{sat}}{1+4\frac{(\Delta\mp k_Lv_{\alpha}- \mu_q(\textbf{r}))^2}{\Gamma^2}}
\end{aligned}
\end{equation}
denotes the saturation parameter for a single beam and atomic transition. 

\leftskip=0.15cm\rightskip=-3cm 
Note that the diffusion is affected by the attenuation of the laser beams and, therefore, is a collective effect that depends on $N$ \cite{17:Temp1, 18:Temp2}. While our model necessarily involves an approximate treatment of diffusion and its dependence on the attenuation, its inclusion is nevertheless important for unstable balanced MOTs \cite{19:noise}. In Sec. \hyperref[sec:3]{III}, we discuss the significance of diffusion in more detail.

\vspace*{-4pt}
\leftskip=2.71cm
\subsubsection{\hspace{-0.21cm}Beam attenuation}\label{sec:a_iii}

\leftskip=0.15cm\rightskip=-3cm
The beam attenuation modifies the properties of all the main physical effects of our model, including itself (see below). Specifically for trapping [see Eq. (\ref{eq:2})], the attenuation's inclusion results in an additional compression known as the shadow force, as illustrated in Fig. \ref{fig:extra}(a). Note that the attenuation in our model is fully justified considering that the experimentally observed optical depths at the instability threshold are typically on the order of 1 (along the $\mathsf{z}$ axis). To describe this effect, we employ the low-saturation-regime assumption, $s\ll1$, under which a given beam intensity decays exponentially as
\begin{equation}
\label{eq:Intensity}
\hspace*{65pt}
\begin{aligned}
I_{\alpha}^{\pm}(\textbf{r},\textbf{v})=I_{\infty}e^{-\mathcal{O}_{\alpha}^{\pm}(\textbf{r},\textbf{v})}
\end{aligned}
\end{equation}

\noindent where $\mathcal{O}_{\alpha}^{\pm}$ is the optical depth of the cloud for a given beam. For, e.g., the positive $\hat{\boldsymbol{\mathsf{z}}}$ directed beam, 
\begin{equation}
\label{eq:OD}
\hspace*{5pt}
\begin{aligned}
\mathcal{O}_{\mathsf{z}}^{+}(\textbf{r},\textbf{v})=&\int_{-\infty}^{z}dz'\,\rho(x,y,z')\\&\times\sum\limits_{q=-,0,+} p_{\mathsf{z},q}^{+}(x,y,z')\sigma_{\mathsf{z},q}^{+}\left(x,y,z',\textbf{v}(x,y,z')\right)
\end{aligned}
\end{equation}
where $\rho$ is the cloud density, and $\textbf{v}(x,y,z')$ refers to the velocity at $(x,y,z')$. For the remaining five MOT beams, analogous expressions of the optical depth can readily be obtained. Note that for the negative $\hat{\boldsymbol{\alpha}}=\hat{\boldsymbol{\mathsf{x}}},\hat{\boldsymbol{\mathsf{y}}},\hat{\boldsymbol{\mathsf{z}}}$ beam, the integral limit is from, respectively, $x,y,z$ to $+\infty$. 

We point out that calculating intensity requires prior knowledge of $I_{tot,q}$, due to it entering into scattering cross sections [Eq. (\ref{eq:3})]. Although the precise determination of $I_{tot,q}$ is numerically demanding, this difficulty can be overcome by implementing an iterative procedure that is presented in Sec. \hyperref[sec:b]{II.B}; thus, the beam cross saturation can be included. Moreover, in Sec. \hyperref[sec:3]{III}, we show how attenuation impacts the instabilities.

\vspace*{-4pt}
\leftskip=2.75cm
\subsubsection{\hspace{-0.21cm}Rescattering force}\label{sec:a_iv}

\begin{figure*}
\vspace*{-112pt}
  \hspace*{-58pt}\includegraphics[scale=0.5]{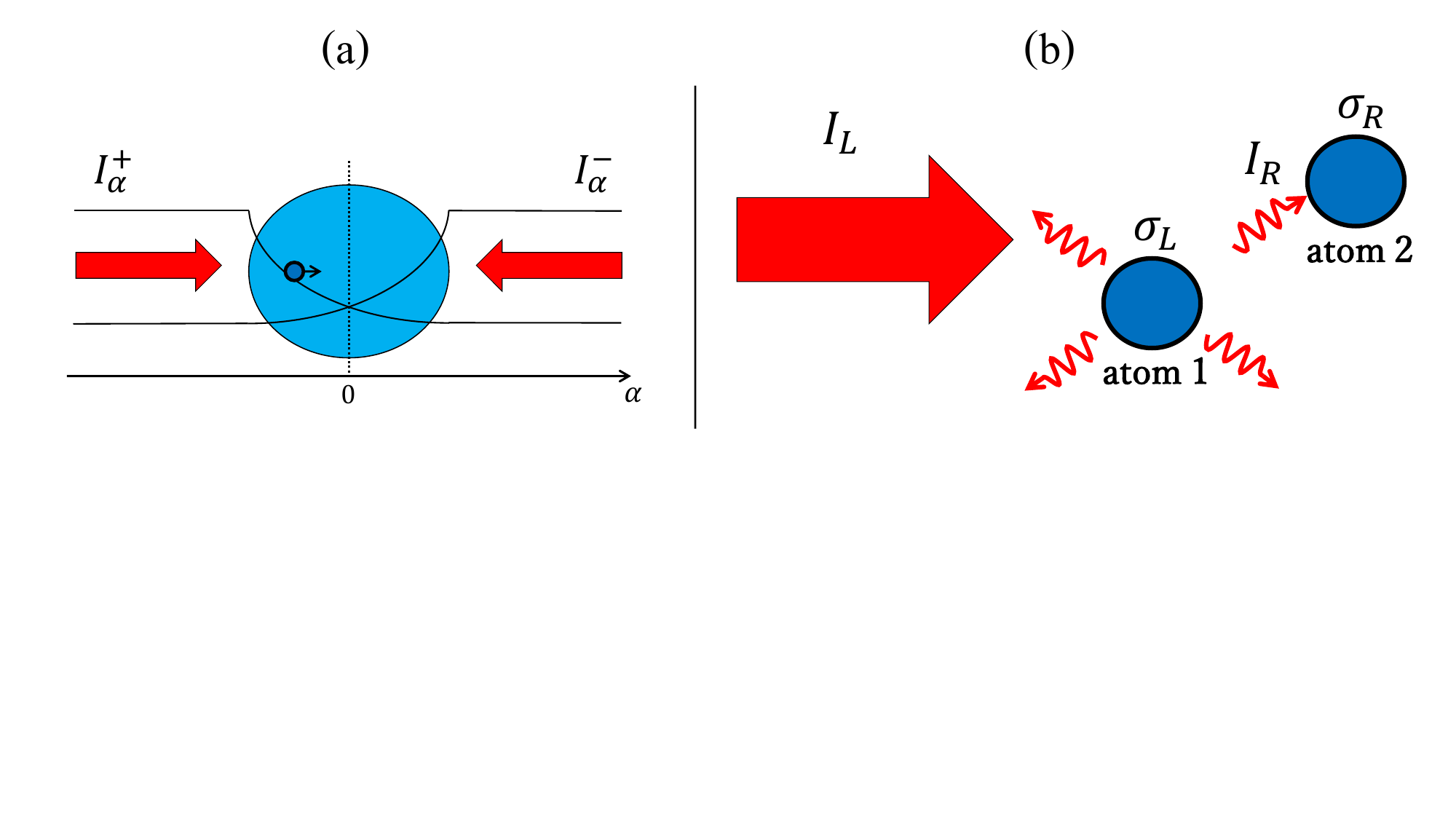}
\vspace*{-143pt}
 \captionsetup{width=1.49\linewidth}
  \caption{(a) Depiction of the mechanism behind the shadow force. As the oppositely directed beams travel through the cloud, their respective intensities $I^{+}_{\alpha}$ and $I^{-}_{\alpha}$ ($\alpha = \mathsf{x},\mathsf{y},\mathsf{z}$) become attenuated, giving rise to an intensity imbalance in the cloud. This imbalance yields an additional compression, i.e., the shadow force. (b) Depiction of the mechanism behind the rescattering force with two 2-level atoms. Atom 1 first scatters the laser light of intensity $I_L$, and atom 2 then rescatters the scattered light, thus experiencing a repulsion. The rescattered light has the intensity $I_R$ determined by the inverse-square law of propagation, such that this repulsion is Coulomb-like. The scattering cross section $\sigma_L$ and the rescattering cross section $\sigma_R$ determine respectively the power scattered and rescattered.}
\label{fig:extra}
\end{figure*}

\leftskip=0.15cm\rightskip=-3cm
To understand the origin of the rescattering force, let us consider the basic situation in Fig. \ref{fig:extra}(b), depicting this force's mechanism with two 2-level atoms. The laser light of intensity $I_L$ is first scattered by atom 1. The scattered light then propagates to atom 2, which rescatters it and thus experiences a repulsive interatomic force. This force

\leftskip=-3cm\rightskip=0.15cm

\noindent has Coulomb-like character, as the rescattered intensity $I_R$ is diminished according to the inverse-square law of propagation. Importantly, the scattering cross section of atom 1, $\sigma_L$, is different from the rescattering cross section of atom 2, $\sigma_R$, due to, e.g., inelastic scattering at atom 1. In the well-known \textit{Wieman} model for the multiple-scattering regime \cite{11:Wieman}, one has $\sigma_R>\sigma_L$, and thus the cloud expands as $N$ is increased; this inequality relies critically on the presence of inelastic scattering in the cloud.

In our model, the rescattering is more complex than described above, and in Fig. \hyperref[fig:rsc_model]{3} we provide an illustration of our employed approximate description. As atom 1 is exposed to all the MOT beams, each of its $\sigma^-$, $\pi$, $\sigma^+$ transitions produces a characteristic radiation pattern of scattered light. At atom 2, each transition of this atom rescatters the scattered light by an amount depending on the fractions that play a role similar to that of the fractions in Eq. (\ref{eq:frac}). 

\vspace*{2pt}
\hspace*{-95pt}\includegraphics[scale=0.46]{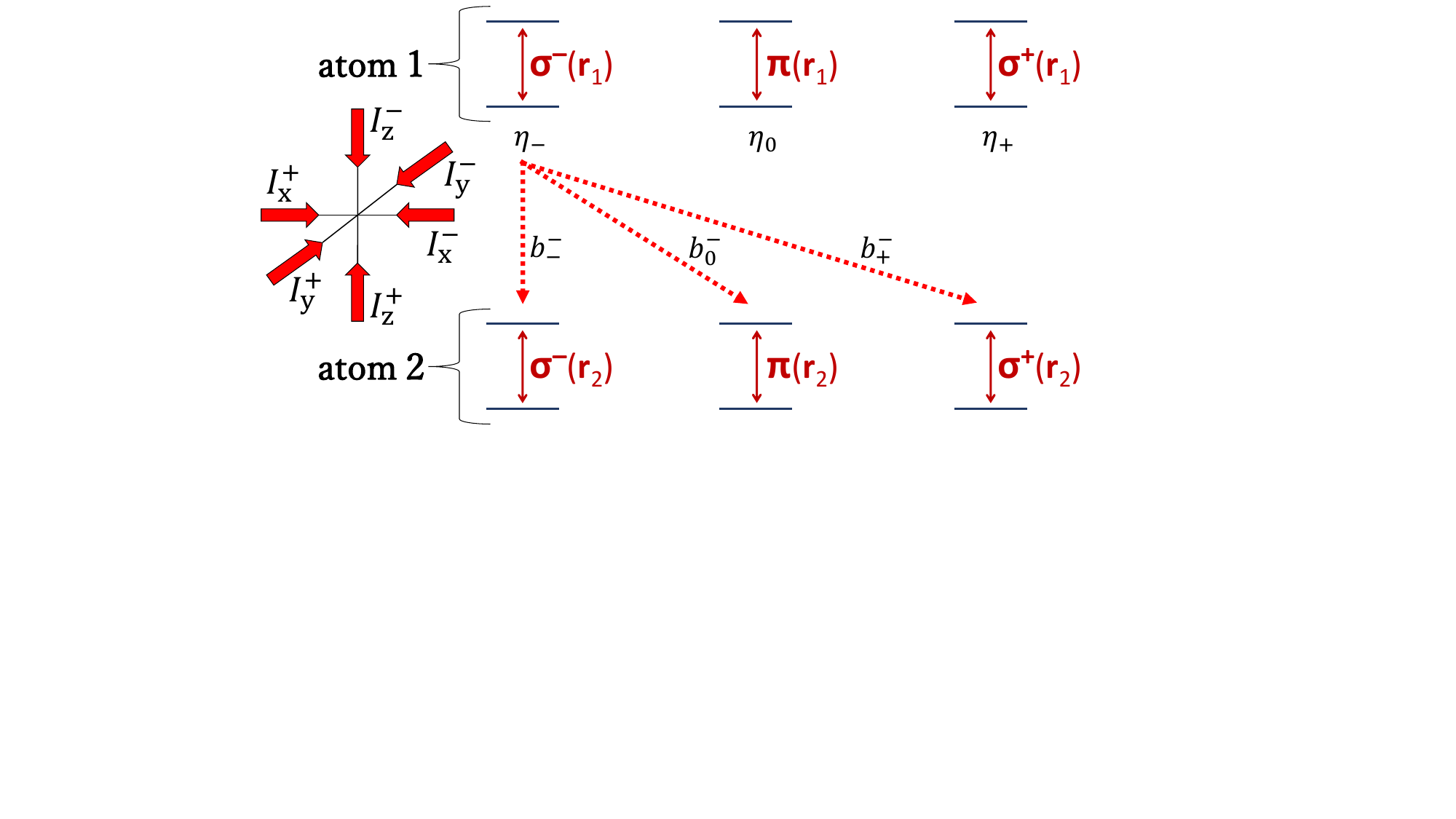}\label{fig:rsc_model}
\vspace*{-126pt}
\begin{footnotesize}
\begin{spacing}{1}
\vspace*{-3pt}\noindent{\textbf{Figure 3:} Illustration of the rescattering in the $F=0 \rightarrow F'=1$ model. Each of the $\sigma^-$, $\pi$, $\sigma^+$ transitions of atom 1 produces a characteristic radiation pattern of light scattered from the total laser field (respectively, $\eta_-$, $\eta_0$, $\eta_+$). The scattered light polarization seen by atom 2 depends on the quantization axis orientations of the two atoms, such that each transition of atom 2 rescatters light in specific fractions (respectively, $b_-^-$, $b_0^-$, $b_+^-$, when considering light from the $\sigma^-$ transition of atom 1).}
\end{spacing}
\end{footnotesize}

\vspace*{8pt}
Under this description, the rescattering force on an atom with the position $\textbf{r}_j$ and the velocity $\textbf{v}_j$, due to surrounding atoms with the positions $\textbf{r}_l$ and the velocities $\textbf{v}_l$, is

\leftskip=0.15cm\rightskip=-3cm 
\noindent
\begin{equation}
\label{eq:F_rsc}
%\hspace*{-65pt}
\hspace*{-85pt}
\begin{aligned}
\textbf{F}_{rsc}(\textbf{r}_j,\textbf{v}_j) = \sum_{l\neq j}\frac{P_R(\textbf{r}_l,\textbf{r}_j,\textbf{v}_l,\textbf{v}_j)}{c}\hat{\textbf{r}}_{l,j}
\end{aligned}
\end{equation} 
where $\hat{\textbf{r}}_{l,j}$ is the unit vector corresponding to $\textbf{r}_{l,j}=\textbf{r}_j-\textbf{r}_l$, which points from atom $l$ to atom $j$, and $P_R$ is the power rescattered by atom $j$ due to atom $l$. It is given by
\begin{equation}
\label{eq:P_R}
\hspace*{7pt}
\begin{aligned}
&P_R(\textbf{r}_l,\textbf{r}_j,\textbf{v}_l,\textbf{v}_j)=\sum_{q''=-,0,+}I_{R,q''}(\textbf{r}_l,\textbf{r}_j,\textbf{v}_l)\sigma_{R,q''}(\textbf{r}_l,\textbf{r}_j,\textbf{v}_l,\textbf{v}_j)\\&=\sum_{q''=-,0,+}\bigg[\sum_{q'=+,0,-}b_{q''}^{q'}(\textbf{r}_l,\textbf{r}_j)I_{S,q'}(\textbf{r}_l,\textbf{r}_j,\textbf{v}_l)\bigg]\sigma_{R,q''}(\textbf{r}_l,\textbf{r}_j,\textbf{v}_l,\textbf{v}_j)
\end{aligned}
\end{equation}
\noindent where $q'$ and $q''$ refer to the 2-level transitions of, respectively, atom $l$ and atom $j$. The sum in the brackets is the intensity $I_{R,q''}$ rescattered by a single 2-level transition of atom $j$. The coefficient $b_{q''}^{q'}$ denotes the fraction of the scattered radiation with the intensity $I_{S,q'}$ that drives its $\sigma^-$, $\pi$, or $\sigma^+$ ($q''=-,0,$ or $+$) transition. Finally, $\sigma_{R,q''}$ is the rescattering cross section for a single 2-level transition of this atom.

The coefficients are given by
\begin{equation}
\hspace*{10pt}
\label{eq:b12}
\begin{aligned}
b_{q''}^{\pm}(\textbf{r}_l,\textbf{r}_j)=\begin{cases}
               \frac{1}{2u}\left[ \left( \displaystyle \prod_{n = l,j} \hat{\textbf{r}}_{l,j} \boldsymbol{\cdot} \hat{\textbf{B}}_n \right) \pm 1 \right]^2\;,\; q'' = +\;(\sigma^+)\\
               \frac{1}{2u}\left[ \left( \displaystyle \prod_{n = l,j} \hat{\textbf{r}}_{l,j} \boldsymbol{\cdot} \hat{\textbf{B}}_n \right) \mp 1 \right]^2\;,\; q'' = -\;(\sigma^-)\\
               1-(b_{+}^{\pm}+b_{-}^{\pm})\quad\quad\quad\quad\quad\quad\,\;\;,\; q'' = 0\;\,\,(\pi)
            \end{cases}
\end{aligned}
\end{equation}
\begin{equation}
\hspace*{10pt}
\label{eq:b3}
\begin{aligned}
b_{q''}^{0}(\textbf{r}_l,\textbf{r}_j)=\begin{cases}
               \frac{1}{2}\left[\hat{\textbf{r}}_{l,j} \boldsymbol{\cdot} \hat{\textbf{B}}_j \right]^2\;\;\;\,\quad\quad\quad\quad\quad\;\;\;,\; q'' = +\;(\sigma^+)\\
               \frac{1}{2}\left[\hat{\textbf{r}}_{l,j} \boldsymbol{\cdot} \hat{\textbf{B}}_j \right]^2\;\;\;\,\quad\quad\quad\quad\quad\;\;\;,\; q'' = -\;(\sigma^-)\\
               1-(b_{+}^{0}+b_{-}^{0})\quad\quad\quad\quad\quad\quad\;\;\,,\; q'' = 0\;\,\,(\pi)
            \end{cases}
\end{aligned}
\end{equation}

\noindent where $\hat{\textbf{B}}_l$ and $\hat{\textbf{B}}_j$ are the unit vectors corresponding to the MOT magnetic fields at, respectively, the atom $l$ and atom $j$ positions, and $u=1+\left(\hat{\textbf{r}}_{l,j} \boldsymbol{\cdot} \hat{\textbf{B}}_l\right)^2$ is a normalization constant common to both $b_{q''}^{+}$ and $b_{q''}^{-}$.

The scattered intensity is given by
\begin{equation}
\hspace*{15pt}
\label{eq:I_S}
\begin{aligned}
 I_{S,q'}(\textbf{r}_l,\textbf{r}_j,\textbf{v}_l)=\eta_{q'}(\textbf{r}_l,\textbf{r}_j)\times\frac{P_{L,tot,q'}(\textbf{r}_l,\textbf{v}_l)}{4\pi |\textbf{r}_{l,j}|^2}
\end{aligned}
\end{equation}

\leftskip=-3cm\rightskip=0.15cm 
\noindent where $P_{L,tot,q'}=\sum_{\alpha=\mathsf{x},\mathsf{y},\mathsf{z}}\left(|\textbf{F}^{+}_{\alpha,q'}| + |\textbf{F}^{-}_{\alpha,q'}|\right)c$ is the total power scattered by a single 2-level transition of atom $l$ [refer to Eq. (\ref{eq:2})], and $\eta_{q'}$ is the corresponding normalized radiation pattern:
\begin{equation}
\hspace*{-90pt}
\label{eq:n_q}
\begin{aligned}
\eta_{q'}(\textbf{r}_l,\textbf{r}_j)=\begin{cases}
               \frac{3}{4}\Big[1+\left(\hat{\textbf{r}}_{l,j}\boldsymbol{\cdot}\hat{\textbf{B}}_l\right)^2\Big]\quad,\quad q' = +\;(\sigma^+)\\
               \frac{3}{4}\Big[1+\left(\hat{\textbf{r}}_{l,j}\boldsymbol{\cdot}\hat{\textbf{B}}_l\right)^2\Big]\quad,\quad q' = -\;(\sigma^-)\\
               \frac{3}{2}\Big[1-\left(\hat{\textbf{r}}_{l,j}\boldsymbol{\cdot}\hat{\textbf{B}}_l\right)^2\Big]\quad,\quad q' = 0\;\,\,(\pi)
            \end{cases}
\end{aligned}
\end{equation}

The rescattering cross section $\sigma_{R,q''}$ is found by evaluating an overlap integral between (a) the emission spectrum that atom $l$ produces when illuminated by the laser field and (b) the absorption spectrum of a single 2-level transition of atom $j$ for the scattered field in the presence of the laser field. Denoting the emission spectrum of atom $l$ by $S_{tot,q''}$ and the absorption spectrum of a single 2-level transition of atom $j$ by $\sigma_{A,q''}$, we have
\begin{equation}
\hspace*{-96pt}
\label{eq:sigma_R}
\begin{aligned}
\sigma_{R,q''}(\textbf{r}_l,\textbf{r}_j,\textbf{v}_l,\textbf{v}_j)=\int d\omega\; S_{tot,q''}(\omega,\textbf{r}_l,\textbf{r}_j,\textbf{v}_l)\sigma_{A,q''}(\omega,\textbf{r}_j,\textbf{v}_j)
\end{aligned}
\end{equation}
\noindent where 

\begin{equation}
\hspace*{-95pt}
\label{eq:Spectrum_tot}
\begin{aligned}
S_{tot,q''}(\omega,\textbf{r}_l,\textbf{r}_j,\textbf{v}_l)=\sum_{q'=+,0,-}w_{q''}^{q'}(\textbf{r}_l,\textbf{r}_j)S_{q'}(\omega,\textbf{r}_l,\textbf{v}_l)
\end{aligned}
\end{equation}
\noindent is composed of the normalized emission spectra $S_{-}$, $S_{0}$, $S_{+}$ for the respective $\sigma^-$, $\pi$, $\sigma^+$ transitions of atom $l$, together with their spectral weight
\begin{equation}
\hspace*{-75pt}
\label{eq:w_q}
\begin{aligned}
w_{q''}^{q'}(\textbf{r}_l,\textbf{r}_j)=\frac{b_{q''}^{q'}(\textbf{r}_l,\textbf{r}_j)\eta_{q'}(\textbf{r}_l,\textbf{r}_j)}{\sum_{q'=+,0,-}b_{q''}^{q'}(\textbf{r}_l,\textbf{r}_j)\eta_{q'}(\textbf{r}_l,\textbf{r}_j)}
\end{aligned}
\end{equation}
where $w_{q''}^{-}+w_{q''}^{0}+w_{q''}^{+}=1$, such that $S_{tot,q''}$ [Eq. (\ref{eq:Spectrum_tot})] is a normalized emission spectrum. 
 
\leftskip=0.15cm\rightskip=-3cm
In Refs. \cite{21:Mollow} and \cite{22:Mollow}, quite general expressions for the emission and absorption spectra for a 2-level atom have respectively been derived. Normalizing these expressions and adapting them to our approximate description, we obtain the expressions for $S_{q'}(\omega,\textbf{r}_l,\textbf{v}_l)=S_{q'}(\omega)$ and $\sigma_{A,q''}(\omega,\textbf{r}_j,\textbf{v}_j)=\sigma_{A,q''}(\omega)$ seen below. We have defined the total Rabi frequency for a single 2-level transition of the $F=0\rightarrow F'=1$ model atom, $\Omega_{tot,q}(\textbf{r},\textbf{v})=\Gamma\sqrt{\frac{I_{tot,q}(\textbf{r},\textbf{v})}{2I_{sat}}}$, and the corresponding detuning incorporating the Zeeman effect, $\Delta_{q}(\textbf{r})=\Delta - \mu_q(\textbf{r})$. Note that $S_{q'}$ is written as a sum of two contributions. The first term, which involves a Dirac delta function, is an elastically scattered spectrum, which is centered at a single frequency specified by $\Delta_{q'}$. The second term is an inelastically scattered spectrum, which, on the other hand, is polychromatic. Because, in the Wieman model, the inelastic scattering is critical for the repulsion to win over the compression, we have been motivated to investigate the impact of elastically and inelastically scattered spectra on the instabilities, and in Sec. \hyperref[sec:3]{III} we see the outcome. The impact of rescattering (as a whole) is also investigated.

Note that the Doppler effect can be included in $\Delta_{q}(\textbf{r})$ in an alternative description, where each of the six MOT beams is separately scattered by the respective three atomic transitions. In such a case, 18 rescattering cross-sections would have to be used, instead of the current 3 [Eq. (\ref{eq:sigma_R})]. Nevertheless, neither of these descriptions is fully justified considering the complexity involved in modeling of the atom's behavior as it is coupled to the interacting field of several beams. In our simulations, we usually have $k_L |v_{\alpha}|\ll \Gamma,|\Delta|$, and thus the Doppler effect can be omitted. However, when deeply in the unstable regime, this is no longer true, and $k_L |v_{\alpha}|$ can become on the order of $\Gamma$. 
\end{multicols}

\small
\vspace*{-12pt}
\par\hspace{10pt}\rule{6cm}{0.4pt}
\begin{equation}
\hspace{-55pt}
\label{eq:S_q}
\vspace*{-12pt}
\begin{aligned}
S_{q'}(\omega)&=\left[\frac{\Gamma^2+4\Delta_{q'}^2}{\Gamma^2+4\Delta_{q'}^2+2\Omega_{tot,q'}^2}\right]\delta(\omega-\Delta_{q'})
\\ &+\frac{\Gamma\Omega_{tot,q'}^2}{2\pi}\left(\frac{(\omega-\Delta_{q'})^2+\frac{1}{2}\Omega_{tot,q'}^2+\Gamma^2}{\Gamma^2\left[\frac{1}{2}\Omega_{tot,q'}^2+\Delta_{q'}^2+\frac{1}{4}\Gamma^2-2(\omega-\Delta_{q'})^2\right]^2 + (\omega-\Delta_{q'})^2\left[\Omega_{tot,q'}^2+\Delta_{q'}^2+\frac{5}{4}\Gamma^2-(\omega-\Delta_{q'})^2\right]^2}\right)
\end{aligned}
\end{equation}
\vspace*{-9pt}

\begin{equation}
\hspace{-51pt}
\label{eq:sigma_A}
\begin{aligned}
\sigma_{A,q''}(\omega)=\frac{\sigma_0\Gamma}{4}&\times\left\{\frac{\Gamma^2+4\Delta_{q''}^2}{\Gamma^2+4\Delta_{q''}^2+2\Omega_{tot,q''}^2}\right\}\\
&\times\left\{\frac{(-i\omega+i\Delta_{q''}+\Gamma)\left(-i\omega+i2\Delta_{q''}+\frac{\Gamma}{2}\right)+\frac{1}{2}i\Omega_{tot,q''}^2(\omega-\Delta_{q''})/\left(i\Delta_{q''}+\frac{\Gamma}{2}\right)}{(-i\omega+i\Delta_{q''}+\Gamma)\left(-i\omega+i2\Delta_{q''}+\frac{\Gamma}{2}\right)\left(-i\omega+\frac{\Gamma}{2}\right)+\Omega_{tot,q''}^2\left(-i\omega+i\Delta_{q''}+\frac{\Gamma}{2}\right)}+\mathsf{c.c.}\right\}
\end{aligned}
\end{equation}
\vspace*{-6pt}

\begin{multicols}{2}\setlength{\columnsep}{2pt}

\normalsize
\leftskip=-3cm\rightskip=0.15cm
We finally discuss the similarities and differences between our kinetic model and that of Ref. \cite{25:Pohl}. In both cases, the same main physical effects except for diffusion are used. In particular, both models include Doppler trapping force, attenuation, rescattering force, with atomic cross-sections possessing complete spatial dependence, thus resulting in all the forces being nonlocal. However, an important difference is that, unlike in Ref. \cite{25:Pohl}, we do not make any assumption of spherical symmetry of the forces. This allows us to observe, e.g., center-of-mass (COM) oscillations of the cloud, which are also seen in experiments \cite{2:MG}. Moreover, we work with the more complex $F=0 \rightarrow F'=1$ system compared to the 2-level system in Ref. \cite{25:Pohl}. This improvement allows us to properly describe, e.g., the anisotropy of the trapping force or, more generally, the features related to the magnetic field and light polarization.

\vspace{6pt}
\leftskip=-0.6cm
\subsection{\hspace{-0.30cm}{Implementation}}\label{sec:b}
\vspace{12pt}

\leftskip=-3cm\rightskip=0.15cm
\noindent The collective system dynamics in the $F=0\rightarrow F'=1$ model (Sec. \hyperref[sec:a]{II.A}) are described by the following collisionless Vlasov-type kinetic equation for the atomic phase-space density $f(\textbf{r},\textbf{v},t)$ \cite{25:Pohl, extra:Vlasov}:
%\vspace*{4pt}
\begin{equation}
\hspace*{-96pt}
\label{eq:Vlasov}
\begin{aligned}
\frac{\partial}{\partial t}f+ \textbf{v}\frac{\partial}{\partial \textbf{r}}f &+ \frac{1}{M}\frac{\partial}{\partial \textbf{v}}\left\{ [ \textbf{F}_{tr}(\textbf{r},\textbf{v}) + \textbf{F}_{rsc,co}(\textbf{r},\textbf{v}) ] f \right\} 
\\&- \frac{1}{M^2} \frac{\partial^2}{\partial \textbf{v}^2} \left[ D(\textbf{r},\textbf{v})f \right] = 0
\end{aligned}
\end{equation}
where $M$ is the atomic mass, $\textbf{F}_{tr}(\textbf{r},\textbf{v})$ is the trapping force given by Eq. (\ref{eq:0}), $D(\textbf{r},\textbf{v})$ is the momentum diffusion coefficient given by Eq. (\ref{eq:5}), and $\textbf{F}_{rsc,co}(\textbf{r},\textbf{v})=\int d\textbf{r}'d\textbf{v}'\,\textbf{F}_{rsc}^{sg}(\textbf{r},\textbf{r}',\textbf{v},\textbf{v}') f(\textbf{r}',\textbf{v}',t)$ is the continuous space analog of $\textbf{F}_{rsc}(\textbf{r},\textbf{v})$ in Eq. (\ref{eq:F_rsc}), where $\textbf{F}_{rsc}^{sg}$ defines the rescattering force due to a single surrounding atom; the positions and velocities depend implicitly on time $t$. Given both the local and nonlocal spatial dependencies of the model's forces, finding direct numerical solution to Eq. (\ref{eq:Vlasov}) is highly demanding. As an equivalent yet more sensible way of solving the dynamics of our system, we use a numerical procedure based on a superparticle \cite{24:super-particle} treatment similar to the one in Ref. \cite{25:Pohl}.

Note that a hydrodynamical description may not be directly comparable to our (kinetic) description. Hydrodynamics typically simplifies the phase space by restricting to, e.g., mean-field velocities. This is not fulfilled especially in the unstable regime. As a consequence, the applicability of such simplified descriptions (see, e.g., the \textit{MOT photon bubble} model \cite{9:fluid}) to the present problem is not obvious.

\leftskip=0.15cm\rightskip=-3cm 
Before getting into detail about the different elements involved in our implementation, we sketch first how it proceeds. At the initial time $t_0$, a Gaussian cloud is generated, composed of superparticles, i.e., particles that  represent collections of regular particles for increasing the simulation speed. Next, the beam intensity attenuation is evaluated at the superparticle positions, with help of our developed tube method \cite{30:Thesis}. Forces acting on each superparticle are then computed, and, using these forces, the Leapfrog algorithm \cite{23:Leapfrog} propagates the cloud in time by one time-step $\delta t$. At the new positions, the attenuation is first evaluated, then the forces are computed, and the algorithm propagates the cloud by one $\delta t$ - this cycle is repeated until the simulation end time $t_{end}$. 

Spatiotemporal instabilities can arise in the simulations, and in Refs. \cite{1:MG, 2:MG} we have demonstrated that we are successful in reproducing experimental behaviors. The 3D nature of our simulations is showcased in Fig. \ref{fig:2} (see also online Supplemental Material \cite{SupplMat}), displaying clouds belonging to different regimes obtained also experimentally (see Ref. \cite{2:MG} for the discussion). Note that in Ref. \cite{25:Pohl}, their kinetic model has been used to achieve simulations of balanced MOT instabilities in quasi-1D.

The exact iterative scheme we employ in the Leapfrog algorithm for updating the superparticle velocity and position in time is as follows:
\newline\newline1. Compute total force $\textbf{F}_{tot}(n)=\textbf{F}_{tot}(\textbf{r}(t_n),\textbf{v}(t_{n-1/2}))$.  \\\\
2. Update velocity $\textbf{v}(t_{n+1/2})=\textbf{v}(t_{n-1/2})+\frac{\textbf{F}_{tot}(n)}{M_{sup}}\delta t$. \\\\
3. Update position $\textbf{r}(t_{n+1})=\textbf{r}(t_{n})+\textbf{v}(t_{n+1/2})\delta t$.
\newline\newline\noindent Here the integer $n=0,1,2, \ldots$ marks the iteration step of fixed size $\delta t$, with $n=0$ used when initiating the scheme, and $M_{sup}=\varepsilon M$ is the superparticle mass, with $\varepsilon=N/N_{sup}$ being a fixed superparticle number $N_{sup}$ scaling (i.e., the number of regular particles represented by one superparticle) that we note also scales $\sigma_0$, so one correspondingly has $\sigma_{sup,0}=\varepsilon\sigma_0$. We have in Refs. \cite{1:MG, 2:MG} used $N_{sup}=7\times10^3$ - a choice that was based on tests for the simulation outcome to be independent of $N_{sup}$; we found that below $N_{sup}=10^3$ the clouds are stable. Observe, in the iterative scheme, that the velocity and position are updated in an interleaved manner, i.e., they \textit{leapfrog} over each other. When the scheme is initiated ($n=0$), the velocity $\textbf{v}(t_{-1/2})$ and the position $\textbf{r}(t_{0})$ are specified. To obtain $\textbf{v}(t_{-1/2})$, we utilize the approximation $\textbf{v}(t_{-1/2})\approx\textbf{v}(t_{0})$, which is assumed to be valid for a small enough $\delta t$. In Refs. \cite{1:MG, 2:MG}, the $\mathsf{x}$, $\mathsf{y}$, $\mathsf{z}$ components of $\textbf{v}(t_{0})$ have been picked to be random between 0 and 0.01 m/s [much\;\;less\;\;than\;\;the\;\;root-mean-square\;\;(RMS)

\leftskip=-3cm\rightskip=0.15cm
\setcounter{figure}{3}
\begin{figure*}
%\vspace*{-287pt}
\vspace*{-275pt}
  \hspace*{-76pt}\includegraphics[scale=0.78]{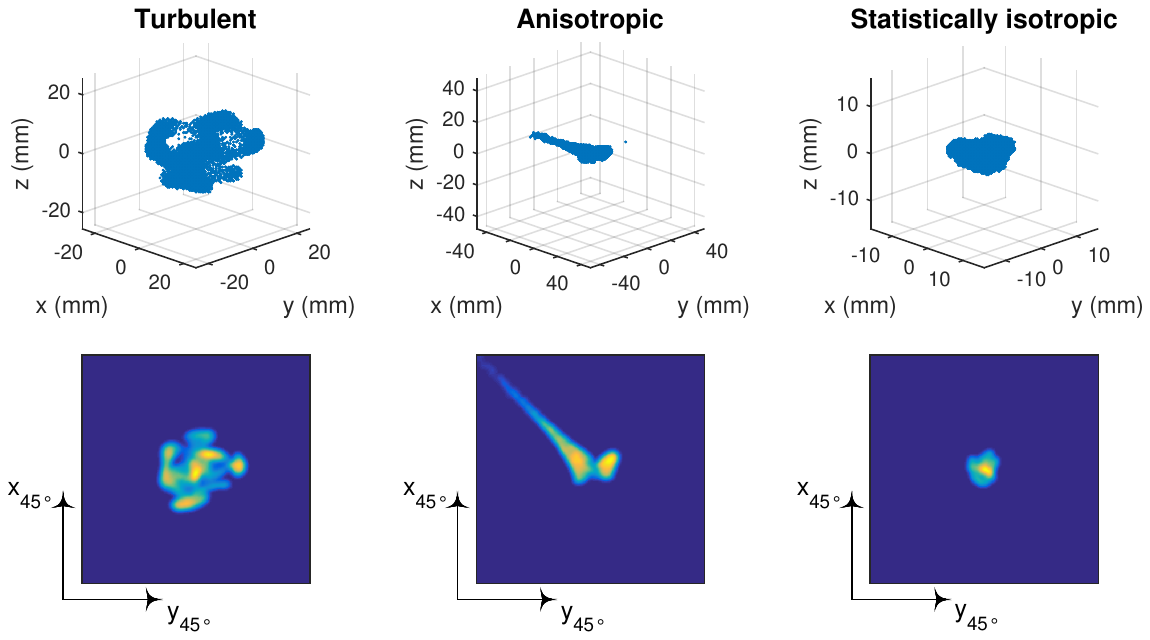}
\vspace*{-206pt}
 \captionsetup{width=1.49\linewidth}
  \caption{Display of 3D images (upper row) and corresponding 2D images (lower row) of simulated clouds belonging to different instability regimes discussed in Ref. \cite{2:MG}. Each image is a single-shot image. In the 3D images, the individual dots are superparticles. The 2D images are the cloud densities integrated along the $\mathsf{z}$ axis and are Gaussian filtered; the diagonals correspond to the directions of two pairs of MOT beams; the field of view is $10\times10$ cm$^2$ for all these images. The $\mathsf{x}$ and $\mathsf{y}$ axes have been rotated by $45^{\circ}$ between the 3D and 2D images. A video version of this figure is available as online Supplemental Material \cite{SupplMat}.}
\label{fig:2}
\end{figure*}

\noindent velocity of Rb-87 at the Doppler temperature]. To pick $\delta t$, a \textit{rule of thumb} is employed, telling that for a given set of MOT parameters we use $\delta t<0.1/\omega_{tr}$, where $\omega_{tr}=\sqrt{\kappa/M}$ is the trap frequency, with $\kappa=\frac{\mu B'}{k_L}\frac{I_{\infty}}{I_{sat}}\frac{-8\hbar k^2_L \Gamma^3 \Delta}{(\Gamma^2 + 6\Gamma^2I_{\infty}/I_{sat}+4{\Delta^2})^2}$ being the trap spring constant [found from the linear expansion of Eq. (\ref{eq:0}) for low velocities and positions near the trap center, with fixed intensity $I_{\infty}$]. In Refs. \cite{1:MG, 2:MG}, we have used the following constants (related to Rb-87 and its D2 line): $M=1.443\times10^{-25}$ kg, $\mu=2\pi\times1.4\times10^6$ Hz/G, $\Gamma=2\pi\times6.07$ MHz, $k_L=\frac{2\pi}{780\times10^{-9}}$ m$^{-1}$, and $I_{sat}=1.67$ mW/cm$^2$.

The total force acting on a superparticle with the position $\textbf{r}_\mathrm{s}=\textbf{r}(t_n)$ and the velocity $\textbf{v}_\mathrm{s}=\textbf{v}(t_{n-1/2})$ is
\begin{equation}
%\hspace*{5pt}
\hspace*{-90pt}
\label{eq:F_tot}
\begin{aligned}
\textbf{F}_{tot}(\textbf{r}_\mathrm{s},\textbf{v}_\mathrm{s})=\varepsilon\textbf{F}_{tr}(\textbf{r}_\mathrm{s},\textbf{v}_\mathrm{s}) + \varepsilon^2\textbf{F}_{rsc}(\textbf{r}_\mathrm{s},\textbf{v}_\mathrm{s}) + \varepsilon\textbf{F}_\mathcal{D}(\textbf{r}_\mathrm{s},\textbf{v}_\mathrm{s})
\end{aligned}
\end{equation}
where $\textbf{F}_\mathcal{D}$ is the stochastic force given by \cite{26:Entropy}

\leftskip=-3cm\rightskip=0.15cm
\begin{equation}
%\hspace*{45pt}
\hspace*{-65pt}
\label{eq:D}
\begin{aligned}
\textbf{F}_\mathcal{D}(\textbf{r}_\mathrm{s},\textbf{v}_\mathrm{s})=\sqrt{\frac{2D(\textbf{r}_\mathrm{s},\textbf{v}_\mathrm{s})}{3\delta t}}\times g(n)\hat{\textbf{r}}_\mathrm{s}
\end{aligned}
\end{equation}
where $D$ is the momentum diffusion coefficient given by Eq. (\ref{eq:5}), and $g(n)$ is a Gaussian white noise with the autocorrelation function equal to $\delta(n)$. We note that, in addition to $\textbf{F}_\mathcal{D}$, the velocity dependence in our total force makes the Leapfrog algorithm no longer time reversible.

In Eq. (\ref{eq:F_tot}), $\textbf{F}_{tr}$ is scaled by $\varepsilon$ as it is written as a sum of terms that contains $\sigma_{\alpha,q}^{\pm}\propto\sigma_0$ [see Eq. (\ref{eq:2})]. $\textbf{F}_{rsc}$ is scaled by $\varepsilon^2$ as it is written as a sum of terms that contains a product of $\sigma_{\alpha,q'}^{\pm}\propto\sigma_0$ and $\sigma_{R,q''}\propto\sigma_0$ [see Eqs.

\leftskip=0.15cm\rightskip=-3cm
\noindent (\ref{eq:P_R}, \ref{eq:I_S}, \ref{eq:sigma_R}, \ref{eq:sigma_A})]. $\textbf{F}_\mathcal{D}$ is scaled by $\varepsilon$ for the following reasons. As the diffusion coefficient $D$ describes the equilibrium between the diffusive heating and Doppler cooling processes, one can write $D\propto\gamma T_{lim}$ \cite{16:Tannoudji}, where the trap friction constant $\gamma$ is scaled by $\varepsilon$ as $\textbf{F}_{tr}$ is scaled likewise, and the limit temperature $T_{lim}$ is scaled by $\varepsilon$ as $T_{lim}\propto M$ according to the equipartition theorem. Moreover, $\textbf{F}_\mathcal{D}$ involves a square root of $D$ [see Eq. (\ref{eq:D})]. Taking everything into account, $\textbf{F}_\mathcal{D}$ is thus scaled by $\varepsilon$.
 
For evaluating the beam intensity attenuation at the superparticle positions, we employ our developed tube method, whose name is attributed to the fact the attenuation of each beam is calculated in rectangular tube-segments parallel to the beam's propagation direction. This method is implemented by extending its 2D illustration detailed in Fig. \hyperref[fig:TubeMethod]{5}. In particular, we numerically generate a fixed grid of points in 3D space and calculate the intensity of each beam at the positions of the grid-points that contain the superparticles, after which the intensity at each superparticle position is found by means of interpolation. The interpolated intensity values are then used in the calculation of the forces [see Eq. (\ref{eq:F_tot})]. Finding a beam intensity at a given grid-point position $\textbf{r}_\mathrm{g}=(x_\mathrm{g},y_\mathrm{g},z_\mathrm{g})$ involves the assumption that a given superparticle at $\textbf{r}_\mathrm{s}=(x_\mathrm{s},y_\mathrm{s},z_\mathrm{s})$ is represented by a Dirac delta function $\delta(\textbf{r}_\mathrm{g}-\textbf{r}_\mathrm{s})$, allowing us to write the cloud density $\rho(\textbf{r}_\mathrm{g})=\sum_\mathrm{s}{\delta(\textbf{r}_\mathrm{g}-\textbf{r}_\mathrm{s})}=\sum_\mathrm{s}{\delta(x_\mathrm{g}-x_\mathrm{s})\delta(y_\mathrm{g}-y_\mathrm{s})\delta(z_\mathrm{g}-z_\mathrm{s})}$. With this assumption, the intensity for, e.g., the positive $\hat{\boldsymbol{\mathsf{z}}}$ directed beam [see Eqs. (\ref{eq:Intensity}, \ref{eq:OD})] is numerically determined from

\end{multicols}
%\vspace*{-15pt}
\vspace*{-35pt}
%\par\hspace{10pt}\rule{6cm}{0.4pt}
\begin{equation}
\label{eq:OD_num}
%\hspace*{-68pt}
\hspace*{-76pt}
\begin{aligned}
\large
I_{\mathsf{z}}^{+}(x_\mathrm{g},y_\mathrm{g},z_\mathrm{g})=I_{\infty}e^{-\mathcal{O}_{\mathsf{z}}^{+}(x_\mathrm{g},y_\mathrm{g},z_\mathrm{g})} \; , \; \mathcal{O}_{\mathsf{z}}^{+}(x_\mathrm{g},y_\mathrm{g},z_\mathrm{g})\approx\frac{\varepsilon}{W^2}\times\sum\limits_{\substack{z_\mathrm{s}<z_\mathrm{g} \\ |x_\mathrm{s}-x_\mathrm{g}|<W/2 \\ |y_\mathrm{s}-y_\mathrm{g}|<W/2}}\left[\sum\limits_{q=-,0,+} p_{\mathsf{z},q}^{+}(x_\mathrm{s},y_\mathrm{s},z_\mathrm{s})\sigma_{\mathsf{z},q}^{+}\left(x_\mathrm{s},y_\mathrm{s},z_\mathrm{s},\textbf{v}_\mathrm{s}(x_\mathrm{s},y_\mathrm{s},z_\mathrm{s})\right) \right]
\end{aligned}
\end{equation}
\vspace*{-10pt}
\hspace{211pt}\noindent\rule{6cm}{0.4pt}
\begin{multicols}{2}\setlength{\columnsep}{2pt}
\leftskip=-3cm\rightskip=0.15cm
\normalsize
\noindent where the superparticle scaling $\varepsilon$ is taken into account, $W$ is a fixed tube-width, and $W^2$ is the corresponding transverse tube-area. From this equation, we observe that the only superparticles that contribute at a given grid-point position are those inside the tube of this point ($|x_\mathrm{s}-x_\mathrm{g}|,|y_\mathrm{s}-y_\mathrm{g}|<W/2$) and positioned before it ($z_\mathrm{s}<z_\mathrm{g}$). The remaining beam intensities are found in an analogous way. The value of $W$ is picked according to our tests' results with uniform and Gaussian clouds, where numerically determined attenuation profiles were checked to converge with corresponding analytically calculated profiles. In the simulations of Refs. \cite{1:MG, 2:MG}, we have used $W=0.15\sigma$, where $\sigma$ is the RMS width of the initial Gaussian cloud ($t_0$), picked to be as close as possible to the RMS radius of the cloud after transient behavior (Fig. 3 in Ref. \cite{1:MG} exemplifies this behavior). We note that the choice of the value of $W$ is a compromise. Indeed, if $W$ is too small compared to $\sigma$, too few superparticles will participate in the determination of the attenuation, thus yielding large spatial fluctuations. In the opposite limit where $W$ is larger than $\sigma$, the spatial dependence of the attenuation will be washed out. In this case, $W$ is expected to limit the size of the structures that can appear in a simulated cloud (such as the ones seen in Fig. \ref{fig:2}).

Lastly, we explain the implementation of the beam cross-saturation effect, which we recall from Sec. \hyperref[sec:a]{II.A} appears because $I_{tot,q}$ enters into scattering cross sections [see Eqs. (\ref{eq:3}, \ref{eq:4})]. To implement this effect, the beam intensities at the grid-point positions are calculated self-consistently. To initiate this calculation, one must specify initial beam intensity values (discussed below), which are used to determine $I_{tot,q}$. With this $I_{tot,q}$, new beam intensities can be found [as in Eq. (\ref{eq:OD_num})]. These new intensities are next used to construct intensities that are equal mixtures of new and last intensities, i.e., we construct [New intensities $+$ Last intensities]$\times 1/2$. The constructed intensities are compared to the last intensities, and this is reiterated until convergent intensities are found. (These convergent intensities are then used in the interpolation of the intensities at the superparticle positions, after which the forces acting on the superparticles are found.) Regarding initial beam intensity values, these are picked to be equal to $I_{\infty}$; any values can, in principle, work, due to the convergent intensities being independent of such a choice. At the second Leapfrog algorithm iteration step, the intensity calculation starts with the convergent beam

\leftskip=0.15cm\rightskip=-3cm
\noindent intensities found in the previous iteration step, and so on for the remaining iteration steps (this increases the simulation speed).

\leftskip=0.15cm\rightskip=-3cm 
%\vspace*{3pt}
\hspace*{-15pt}\includegraphics[scale=0.52]{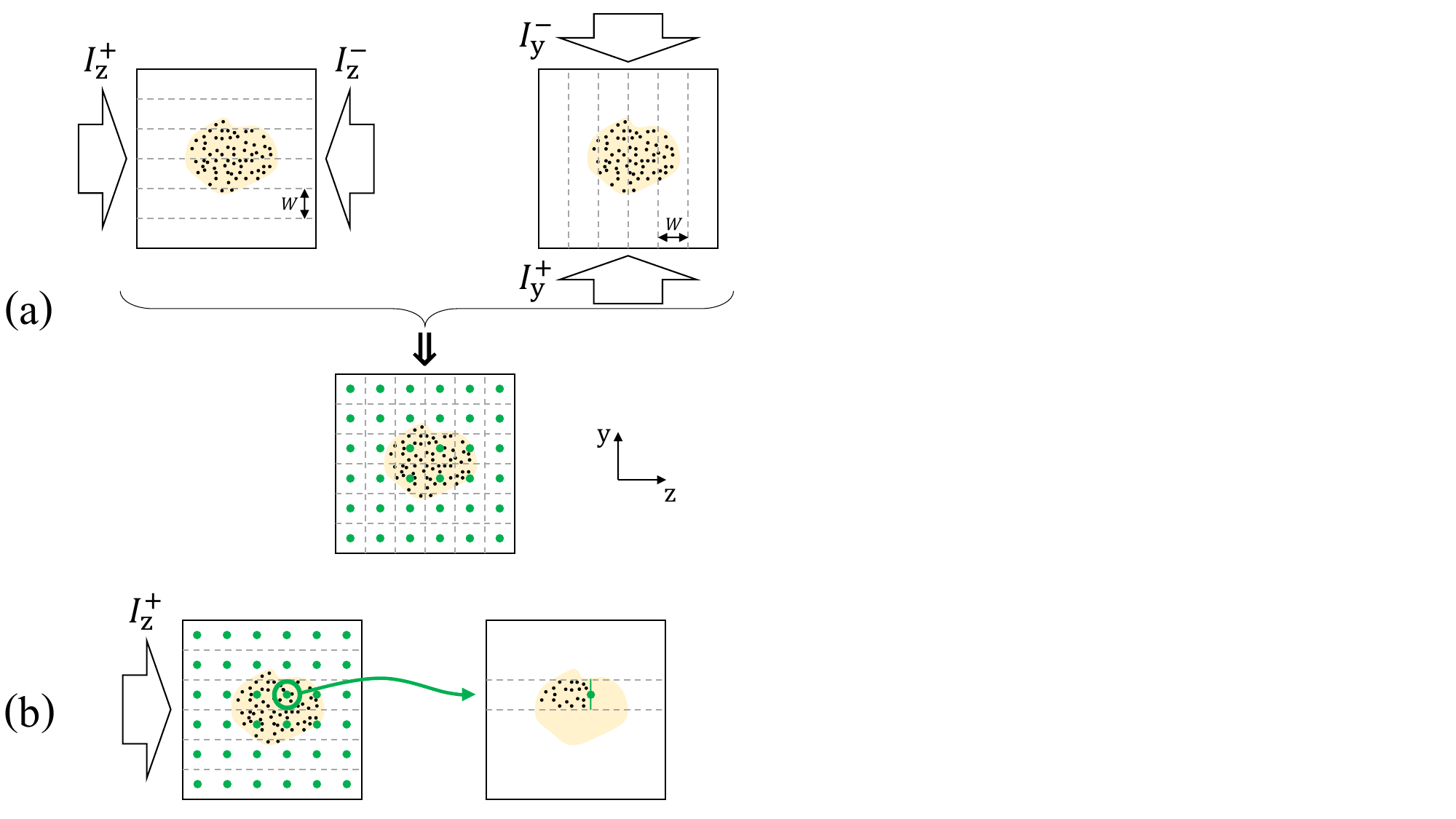}\label{fig:TubeMethod}
\begin{footnotesize}
\begin{spacing}{1}
\vspace*{-7pt}\noindent{\textbf{Figure 5:} (a) First and (b) second part of a 2D illustration of the numerical method called the tube method. (a) The upper two drawings display the same cloud composed of superparticles (small dots), and it is imagined that each beam is segmented into rectangular tubes parallel to the beam's propagation direction. During the cloud's evolution, the tubes remain at fixed positions, with their width $W$ being of a set size. The lower drawing displays the same cloud, with the added large dots indicating the positions of the points where the intensity of each beam is calculated first. These dots are placed at grid-point positions located through the center of the tubes. (b) In the calculation of, e.g., the $+\hat{\boldsymbol{\mathsf{z}}}$ directed beam's intensity at the position of a given large dot, the superparticles that contribute are those inside the tube of this dot and positioned before it. Once the intensities of each beam at the positions of the large dots are calculated, the intensities at the positions of the superparticles are found by means of interpolation.}
\end{spacing}
\end{footnotesize}
%\vspace*{8pt}

%\leftskip=0.32cm
%\vspace{-15pt}
%\section{\hspace{-0.38cm}{Impact of different effects on \\ \hspace*{25pt}{the instabilities} }}\label{sec:3}
%\vspace{-3pt}

\leftskip=-2.7cm
\section{\hspace{-0.38cm}{Impact of different effects on \\ \hspace*{25pt}{the instabilities} }}\label{sec:3}
\vspace{-3pt}

%\leftskip=0.15cm\rightskip=-3cm 
\leftskip=-3cm\rightskip=0.15cm
The simulations offer a better understanding of the complex instability mechanism by varying the magnitude of the different effects. In the following, we illustrate this with preliminary tests on the impact of diffusion, attenuation, and rescattering, as well as of elastic versus inelastic scattering. In doing so, we look at how the detuning at threshold, $\Delta_{thr}$, is affected. Indeed, the detuning is the most sensitive parameter for determining the state (stable or unstable) of a MOT \cite{3:stellar}.

\leftskip=-3cm\rightskip=0.15cm
Let us begin by investigating how $\Delta_{thr}$ is affected as the stochastic force $\textbf{F}_\mathcal{D}$ [Eq. (\ref{eq:D})] is scaled by a constant factor $d$. We use $d = 0, 0.5, 2, 5$ and concentrate on the case with $B' = 3$ G/cm. We display, in Fig. \hyperref[fig:6]{6}, the outcome of this investigation. As can be seen, with no diffu-

\leftskip=-3cm\rightskip=0.15cm
\vspace*{-151pt}
\hspace*{-80pt}\includegraphics[scale=0.63]{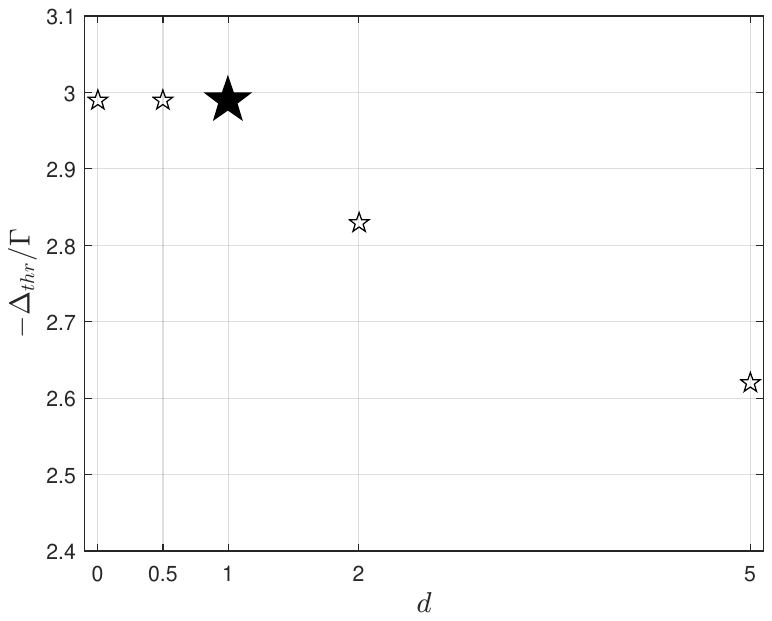}\label{fig:6}
\vspace*{-160pt}
\begin{footnotesize}
\begin{spacing}{1}
\noindent{\textbf{Figure 6:} Investigation on how the threshold detuning $\Delta_{thr}$ for the magnetic field gradient $B'=3$ G/cm is affected as the stochastic force $\textbf{F}_\mathcal{D}$ [Eq. (\ref{eq:D})] is scaled by a constant factor $d$. The open stars are the test results, and the solid star is the result of the ordinary simulation (from Fig. 6 in Ref. \cite{1:MG}).}
\end{spacing}
\end{footnotesize}
\vspace{9pt}

\noindent sion ($d=0$), $\Delta_{thr}$ is the same as in the ordinary simulation ($d=1$; $-\Delta_{thr}/\Gamma=2.99$). Importantly, because the instabilities can exist without the diffusion, this tells us that it is not essential for triggering them. We note that such a conclusion stands in agreement with the results of Ref.\;\;\cite{25:Pohl}, where diffusion was neglected. Our studied instabilities are thus different from retroreflected MOT instabilities of the \textit{stochastic} type, where noise is necessary \cite{27:RR2}. With an increased diffusion ($d>1$), $\Delta_{thr}$ is seen to shift closer to the resonance, with the change being ${\sim}0.4\,\Gamma$ per half a decade. A smaller $-\Delta_{thr}$ means a reduced instability range and, therefore, a weakened instability mechanism. This may not be surprising conside-

\leftskip=0.15cm\rightskip=-3cm
\noindent ring that, in a feedback system, the phase of the feedback is critical in determining whether the system is stable or unstable \cite{extra:Feedback}; increasing the diffusion may be regarded as destroying the phase relationship, thus preventing the unstable regime from being entered. In conclusion, the diffusion is not an essential effect but, otherwise, suppresses the instability mechanism if it becomes great enough.

Next, we continue with our investigation on whether the instabilities persist when either attenuation is removed (i.e., the beam intensity is constant) or rescattering is removed, for $B'=3$ G/cm at different $\Delta$ values in the previous range of simulated unstable and stable clouds \cite{2:MG}. Note that while the former case is experimentally relevant (for relatively low atom number and/or weak magnetic field gradient), the latter one is not, as one cannot have large attenuation without multiple scattering. Without attenuation, we find for all the explored parameter range that the MOT is stable. When attenuation is present but rescattering is turned off, we observe small clouds with temporal fluctuations in their COM positions, but relatively stable RMS radii. The COM fluctuations are in this particular case attributed to diffusion. These tests thus seem to show that both attenuation and rescattering are necessary to reach the unstable regime. This is consistent with previous models \cite{3:stellar, 25:Pohl} and supports the view that the shadow force (produced by attenuation) takes part in a feedback mechanism, where this force works against the cloud expansion due to the rescattering force in order to produce unstable motion in a balanced MOT. The beam attenuation alone can be noted to be critical for instabilities in a retroreflected MOT \cite{27:RR2, 28:RR3} as well as in the case of collective phenomena that parameter-modulated MOT instabilities can exhibit \cite{29:PM2}. Moreover, Ref. \cite{30:Thesis} has recently identified both attenuation and rescattering as necessary effects for a spatiotemporal instability in a misaligned MOT, using a slightly modified version of our simulation model.

Finally, we investigate the impact of elastically versus inelastically scattered light on the instabilities. Here we use the notations $\sigma_{el,q}$ and $\sigma_{inel,q}$ in denoting the parts of $\sigma_{R,q}$ [Eq. (\ref{eq:sigma_R}), with $q''=q$] that result from the contribution of, respectively, elastically and inelastically scattered spectra; they satisfy $\sigma_{R,q} = \sigma_{el,q} + \sigma_{inel,q}$. In Fig. \hyperref[fig:7]{7}, we display how $\Delta_{thr}$ is affected for $B'=3$ G/cm after removing either part. Importantly, the fact the instabilities still are obtained indicates that none of these parts alone is necessary. This is surprising considering the (before-mentioned) Wieman model prediction that the cloud expansion relies critically on the presence of inelastic scattering in the cloud. The thresholds are seen to be shifted further away from the resonance compared to the ordinary simulation result ($-\Delta_{thr}/\Gamma=2.99$), by\;\;${\sim}2.5$ times

\leftskip=-3cm\rightskip=0.15cm
\noindent ($-\Delta_{thr}/\Gamma=7.22$ for $\sigma_{el,q}=0$ and $-\Delta_{thr}/\Gamma=8.06$ for $\sigma_{inel,q}=0$). The shift to larger absolute values is correlated with the fact that we observe the simulated clouds to become smaller (by ${\sim}1.5$ times, with the $\sigma_{inel,q}=0$ case being slightly smaller in cloud size), which is consistent with the Wieman model (there is less rescattering). With this decrease, one is led to an increase in the optical depth and thus the shadow force. On the opposite hand, we find that artificially increasing $\sigma_{R,q}$, and consequently the size, shifts the threshold closer to the resonance. This result makes perfect sense considering the second investigation's finding (on attenuation), as for a vanishing shadow force the instabilities should disappear. 

\vspace*{-150pt}
\hspace*{-80pt}\includegraphics[scale=0.63]{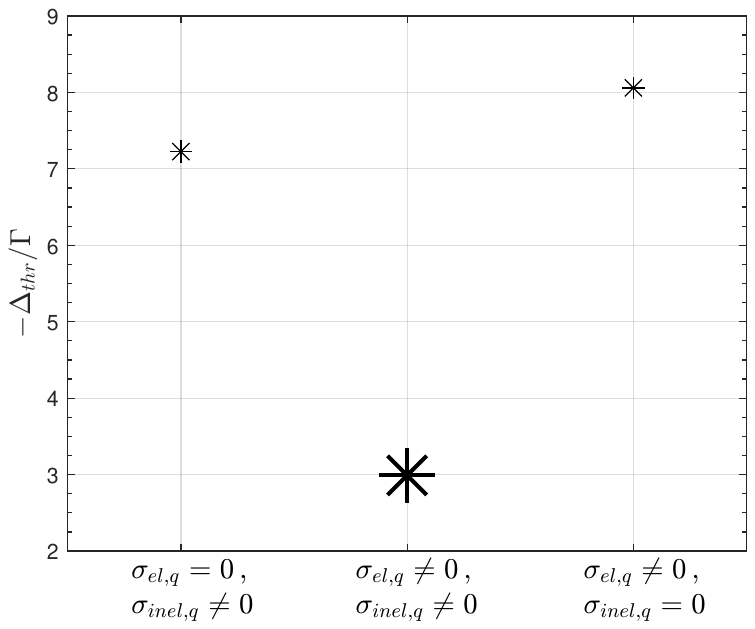}\label{fig:7}
\vspace*{-162pt}
\begin{footnotesize}
\begin{spacing}{1}
\noindent{\textbf{Figure 7:} Investigation on how the threshold detuning $\Delta_{thr}$ for the magnetic field gradient $B'=3$ G/cm is affected after removing either $\sigma_{el,q}$ or $\sigma_{inel,q}$, being the parts of rescattering cross section that result from the contribution of, respectively, the elastically and inelastically scattered spectra. The small asterisks are the test results, and the large asterisk is the result of the ordinary simulation (from Fig. 6 in Ref. \cite{1:MG}).}
\end{spacing}
\end{footnotesize}
\vspace{9pt}

To summarize, both attenuation and rescattering seem to be necessary for generating instabilities in a balanced MOT, and their mechanism is strengthened when the shadow force (produced by attenuation) gets larger compared to the rescattering force. Also, great enough diffusion suppresses this mechanism, and neither elastically nor inelastically scattered light alone is critical for the generation.

\leftskip=-0.36cm
\vspace*{-7pt}
\section{\hspace{-0.38cm}{Conclusion}}\label{sec:4}

\leftskip=-3cm\rightskip=0.15cm
In this paper, we presented a numerical tool for studying MOT instabilities in a full-blown 3D environment. It has been successfully employed to predict instability thresholds \cite{1:MG} and various unstable regimes \cite{2:MG} in a large balanced MOT. These simulations can be used in the future

\leftskip=0.15cm\rightskip=-3cm
 \noindent to investigate features that are challenging to access experimentally, such as, e.g., velocity fields and 3D density distributions. Other lines of research include the analysis of the cloud dynamics in terms of turbulence, to compare with recent experimental observations \cite{extra:Nazarenko}. With some minor modifications, our model is applicable to different MOT configurations, as recently exemplified for a misaligned MOT \cite{30:Thesis}. In the case of a retroreflected MOT, the added value of the spatial information could be strongly beneficial for studying the observed spatiotemporal instabilities \cite{32:RR4}. Improving the understanding of MOT instabilities can effectively continue through investigations on how these are impacted by simulation model's physical effects, and broader perspectives can be opened up by refining the descriptions of currently included effects and/or by adding new ones (e.g., dipole forces, higher-order rescattering).

\leftskip=2.6cm
%\vspace{-5pt}
\vspace{10pt}
\section*{\hspace{-0.55cm}{ Acknowledgments }}
\vspace{3pt}

\leftskip=0.15cm\rightskip=-3cm
Part of this work was performed in the framework of the European training network ColOpt, which is funded by the European Union (EU) Horizon 2020 programme under the Marie Sklodowska-Curie action (Grant No. 721465), and of the European project ANDLICA, ERC advanced grant agreement (Grant No. 832219). It has also been supported by the Danish National Research Foundation through a Niels Bohr Professorship to T.P. and through the Center of Excellence "CCQ" (Grant No. DNRF156). 
\newline\newline
\noindent\textbf{Data availability}: The data supporting this work are available from the corresponding author upon reasonable request.
\vspace*{10pt}
\par\hspace{56pt}\rule{4cm}{0.6pt}
\hspace{-88pt}\rule{2cm}{2pt}
\hspace{-74pt}\rule{3cm}{1.1pt}
\begingroup
\renewcommand{\section}[2]{}

\let\OLDthebibliography\thebibliography
\renewcommand\thebibliography[1]{
  \OLDthebibliography{#1}
  \setlength{\parskip}{4pt}
  \setlength{\itemsep}{1pt plus 0.3ex}
}

\endgroup

\end{multicols}

\end{document}